\documentclass[conference]{IEEEtran}
\IEEEoverridecommandlockouts
\usepackage{cite}
\usepackage{amsmath,amssymb,amsfonts}
\usepackage{algorithmic}
\usepackage{graphicx}
\usepackage{textcomp}
\usepackage{xcolor}
\usepackage{hyperref}
\usepackage{enumitem}
\usepackage{array}
\usepackage{multirow}
\usepackage{makecell}
\usepackage{adjustbox}
\usepackage{listings}
\usepackage{tcolorbox}
\tcbuselibrary{listingsutf8}
\usepackage{stfloats}

\usepackage{xcolor}
\usepackage{color,soul}

\newcolumntype{L}[1]{>{\raggedright\arraybackslash}p{#1}} 

\def\BibTeX{{\rm B\kern-.05em{\sc i\kern-.025em b}\kern-.08em
    T\kern-.1667em\lower.7ex\hbox{E}\kern-.125emX}}
    
\hypersetup{
    pdfauthor={},          
    pdftitle={Conference Paper Title},  
    pdfkeywords={}
}

\begin{document}

\title{SPADE: Enhancing Adaptive Cyber Deception Strategies with Generative AI and Structured Prompt Engineering}

\author{\IEEEauthorblockN{Shihab Ahmed\IEEEauthorrefmark{1}, A B M Mohaimenur Rahman\IEEEauthorrefmark{2}, Md Morshed Alam\IEEEauthorrefmark{3}, Md Sajidul Islam Sajid\IEEEauthorrefmark{1}}
	\IEEEauthorblockA{
    \IEEEauthorrefmark{1}Department of Computer and Information Sciences, Towson University, Towson, Maryland, USA\\
	\IEEEauthorrefmark{2}Department of Software and Information Systems, UNC Charlotte, Charlotte, North Carolina, USA\\
    \IEEEauthorrefmark{3}School of Cybersecurity, Old Dominion University, Norfolk, Virginia, USA\\
	sahmed23@students.towson.edu, arahman3@charlotte.edu, m2alam@odu.edu, msajid@towson.edu}
} 

\maketitle



\begin{abstract} The rapid evolution of modern malware presents significant challenges to the development of effective defense mechanisms. Traditional cyber deception techniques often rely on static or manually configured parameters, limiting their adaptability to dynamic and sophisticated threats. This study leverages Generative AI (GenAI) models to automate the creation of adaptive cyber deception ploys, focusing on structured prompt engineering (PE) to enhance relevance, actionability, and deployability. We introduce a systematic framework (SPADE) to address inherent challenges large language models (LLMs) pose to adaptive deceptions, including generalized outputs, ambiguity, under-utilization of contextual information, and scalability constraints. Evaluations across diverse malware scenarios using metrics such as Recall, Exact Match (EM), BLEU Score, and expert quality assessments identified ChatGPT-4o as the top performer. Additionally, it achieved high engagement (93\%) and accuracy (96\%) with minimal refinements. Gemini and ChatGPT-4o Mini demonstrated competitive performance, with Llama3.2 showing promise despite requiring further optimization. These findings highlight the transformative potential of GenAI in automating scalable, adaptive deception strategies and underscore the critical role of structured PE in advancing real-world cybersecurity applications.
\end{abstract}


\section{Introduction}

The increasing sophistication of modern malware presents a critical challenge to cybersecurity defenses. Advanced tactics, techniques, and procedures (TTPs) such as polymorphism, obfuscation, and targeted exploitation allow attackers to evade traditional detection systems, rendering conventional defenses inadequate. Cyber deception has emerged as a powerful complementary strategy, leveraging techniques such as honeypots, honey tokens, and decoy systems to mislead attackers, delay their progress, and gather intelligence on their behavior \cite{virtual_honeypot_framework, virtual_honeypot_botnet_tracking, honeyproxy_design_implementation}. While effective in some contexts, traditional deception techniques often rely on manual configurations and static parameters that lack the adaptability required to counter rapidly evolving threats \cite{detecting_shadow_honeypots, urias2017technologies, chimera_autonomous_planning}.

Recent advancements in automation have introduced systems capable of dynamically orchestrating deception strategies based on malware analysis \cite{soda_system_for, chimera_autonomous_planning, symbsoda_configurable_verifiable}. These systems represent a significant step forward by analyzing malware behaviors to determine which pre-set rules or static deception ploys to deploy at runtime. However, their primary limitation lies in the reliance on a finite set of manually crafted ploys, which restricts their ability to adapt to diverse or novel threat scenarios. The process of creating these ploys is labor-intensive and provides limited coverage, leaving gaps in addressing the full spectrum of sophisticated malware behaviors. Generative AI (GenAI) presents a transformative alternative by automating the creation of adaptive, context-aware deception ploys. Capable of generating scalable and diverse strategies tailored to specific malware tactics, GenAI reduces manual effort while significantly enhancing runtime adaptability and real-world applicability.

Despite the promise of GenAI, its application in cyber deception remains relatively unexplored and narrowly implemented. Most existing studies \cite{act_as_honeytoken, chatbots_honeypot_world, engaging_attackers_highly, llm_in_shell} focus on generating honeypots and honey tokens, often relying on predefined parameters that limit adaptability. While these efforts lay a foundational framework for GenAI-driven deception, they largely overlook the broader potential for dynamic, adaptive strategies. Specifically, they fail to leverage GenAI's capability to enable context-aware orchestration, multi-layered deception tactics, and deception ploys tailored in real-time to evolving malware behaviors. Expanding beyond static resource generation to encompass more sophisticated, adaptive mechanisms could vastly improve the scalability and efficacy of cyber deception. This research bridges this gap by empirically analyzing the performance of GenAI models—including ChatGPT (Omni \& Mini), Gemini, and Llama3.2—in autonomously generating diverse, adaptive, and realistic deception strategies.

A key enabler of GenAI’s potential in cyber deception is structured PE, which designs specific input prompts to guide GenAI models in producing precise, actionable, and deployable outputs. Recent research demonstrates that well-crafted prompts significantly enhance the relevance and effectiveness of GenAI outputs \cite{prompt_pattern_catalog, giray2023prompt}. PE ensures that the generated deception strategies align with the operational context and threat landscape while reducing the need for manual intervention. This work introduces a systematic framework, \textbf{S}tructured \textbf{P}rompting for \textbf{A}daptive \textbf{D}eception \textbf{E}ngineering (\textbf{SPADE}), to standardize and optimize the prompting process. SPADE enables GenAI models to consistently produce high-quality, deployable deception strategies aligned with real-world cybersecurity scenarios. Specifically, this study makes the following key contributions:

\begin{itemize}[noitemsep,topsep=1pt, leftmargin=9pt] 
\item \textbf{Comprehensive Empirical Evaluation of GenAI for Cyber Deception:} We evaluate the performance of multiple GenAI models, including ChatGPT-4o, ChatGPT-4o Mini, Gemini, and Llama3.2, in autonomously generating diverse deception ploys such as honeyfiles, patches, and API hooks. This analysis provides critical insights into the models’ effectiveness, scalability, and applicability for real-world cyber deception.

\item \textbf{Expansion of Automated Deception Ploys Beyond Traditional Honeypots:} This research explores the broader spectrum of deception ploys that GenAI can produce, moving beyond static honeypots and honey tokens. By automating a diverse set of ploys, this study demonstrates how GenAI can generate more adaptive and layered deception strategies, equipping defenders with a more versatile toolkit to counter evolving cyber threats.

\item \textbf{Systematic Framework for Prompt Engineering:} Recognizing the importance of prompt design, we propose SPADE, a structured framework for adaptive deception engineering. SPADE standardizes the PE process across multiple GenAI models, ensuring consistent and deployable outputs while reducing reliance on manual adjustments.
\end{itemize}


The paper is organized as follows: Section \ref{RW} reviews related work; Section \ref{meth} details the methodology, including the structured prompt engineering framework and system workflow; Section \ref{eval} presents the evaluation; and Section \ref{con} concludes with findings and implications.

\section{Related Works}\label{RW}

Cyber deception has been a cornerstone of cybersecurity, designed to mislead attackers and gather intelligence. Traditional approaches, such as honeypots, honeytokens, and honeynets, create controlled environments to study malicious behavior without risking real systems \cite{virtual_honeypot_framework, virtual_honeypot_botnet_tracking, honeyproxy_design_implementation, detecting_shadow_honeypots}. These techniques effectively deflect attackers, waste their time, and provide insights into their tactics \cite{enhancing_cyber_attribution, unveiling_attack_patterns}. However, static solutions lack adaptability and scalability, leading to the development of dynamic deception techniques. For instance, \cite{ahead_active_defense} proposed integrating deception directly into production systems for realistic interactions and enhanced adversary monitoring. Advanced systems \cite{symbsoda_configurable_verifiable, soda_system_for, chimera_autonomous_planning, sajid2020dodgetron, optimal_honeypot_allocation} leverage the MITRE ATT\&CK framework to dynamically design deception ploys, improving real-time threat mitigation and adaptability.

Recent efforts have explored automating the generation of deception artifacts to address these limitations. High-fidelity fake documents with embedded monitoring mechanisms have been proposed \cite{cognitive_deception_model, controllable_fake_document}, while \cite{generation_deployment_honeytoken} introduced a machine learning framework for deploying honeytokens in relational databases. These studies emphasize the importance of realistic and adaptable deception to counter increasingly sophisticated threats. Generative AI (GenAI) extends this adaptability, automating the creation of honeypots, honeytokens, and fake documents while reducing manual, error-prone processes. By leveraging LLMs, GenAI enables the dynamic generation of context-aware deception artifacts tailored to evolving threats. Furthermore, GenAI can analyze threat intelligence reports to assist deception engineers in crafting effective ploys, streamlining development and enhancing scalability.

Recent works demonstrate the versatility of GenAI in cyber deception. Studies have explored dynamic honeypot systems mimicking command-line environments \cite{chatbots_honeypot_world}, interactive honeypots using ChatGPT APIs \cite{engaging_attackers_highly}, and adaptive systems like \textit{shelLM} for generating realistic responses to attacker inputs \cite{llm_in_shell}. Generative models such as Tabular Masked Transformer (TabMT) have been employed to generate realistic NetFlow traffic for deception \cite{advanced_cyber_deception_framework}, while others focus on low-risk LLM-driven honeypots that simulate terminal environments \cite{on_designing_low_risk}. Beyond honeypots, applications like automated phishing countermeasures \cite{scamming_the_scammers} and realistic query responses for MySQL honeypots \cite{mysqlplot_llm_based_honeypot} highlight GenAI’s adaptability across various domains.

Despite these advancements, many existing studies are confined to generating static resources like honeypots and honeytokens with predefined configurations. They do not fully leverage GenAI's capacity for adaptive, multi-layered deception strategies or real-time orchestration tailored to specific malware behaviors. This work addresses this gap by conducting an empirical evaluation of GenAI models—including ChatGPT (Omni \& Mini), Gemini, and Llama3.2—focusing on their ability to autonomously generate diverse and realistic deception ploys. By introducing a unified framework for structured PE and adaptive deception, this study advances the scalability, precision, and applicability of GenAI-driven cyber deception strategies.

\begin{figure*}[!ht]
    \centering
    \includegraphics[width=\textwidth] {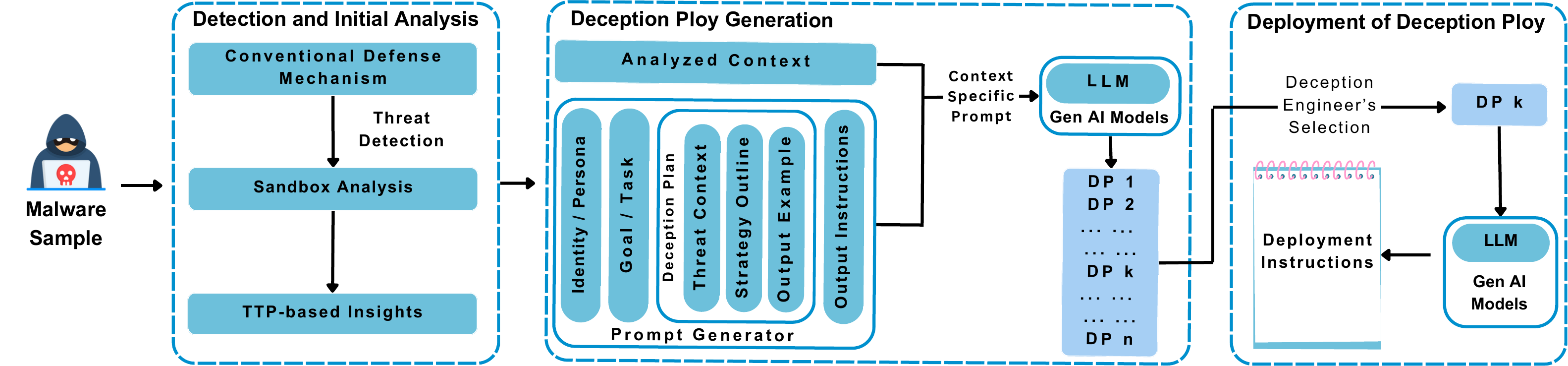}
    \caption{System Architecture of SPADE (DP stands for Deception Ploy).}
    \label{fig:system_architecture}
\end{figure*}

\begin{table*}
\caption{Systematic Prompt Structuring for Adaptive Malware Deception.}
\centering
\scriptsize
\renewcommand{\arraystretch}{1.5} 
\scalebox{1.1}{
\begin{tabular}{|l|p{1.8cm}|L{4.0cm}|L{4.0cm}|L{4.1cm}|} 
\hline
\multicolumn{2}{|l|}{\textbf{Component}} &
  \textbf{Motivation} &
  \textbf{Specification} &
  \textbf{Example} \\ \hline
\multicolumn{2}{|l|}{Identity/Persona/Role} &
  Ensures alignment with domain-specific tasks by assigning the model relevant persona or role. &
  Specifies the role or persona the GenAI should assume (e.g., security analyst or attacker). &
 ``Act as a cybersecurity expert tasked with generating a deception strategy for ransomware X." \\ \hline
\multicolumn{2}{|l|}{Goal/Task} &
  Prevents ambiguity by explicitly defining the desired outcome of the prompt. &
  Clearly states the task GenAI should perform, ensuring the output aligns with operational goals. &
  ``Generate a honeytoken to engage ransomware targeting credential-based exploits." \\ \hline
\multicolumn{1}{|c|}{\rotatebox{90}{\hspace{-20mm}\textbf{Deception Plan}}} &
  \raggedright\arraybackslash Threat Context \newline (Meaningful Context) &
  Provides GenAI with malware-specific behavior and environmental insights for targeted outputs. &
  Describes sandbox results, observed TTPs, and attack patterns to embed malware context. &
  ``The ransomware encrypts files in directories named 'docs.' Generate deception to exploit this." \\ \cline{2-5} 
\multicolumn{1}{|l|}{} &
  \raggedright\arraybackslash Strategy Outline \newline (Dos \& Don’ts) &
  Guides the model to create operationally feasible and resource-efficient deception strategies. &
  Includes operational constraints and specific tactics GenAI should use or avoid. &
  ``Avoid high-resource solutions; focus on lightweight honeyfiles and decoy tokens." \\ \cline{2-5} 
\multicolumn{1}{|l|}{} &
  \raggedright\arraybackslash Output Example/ Guidance (Few-\newline Shot Prompting) &
  Improves consistency by guiding GenAI with template outputs or examples. &
  Supplies examples of ideal outputs or templates to align generated results with practical requirements. &
  ``Include a sample JSON for a honeytoken or a pseudocode snippet for an API hook. \\ \hline
\multicolumn{2}{|l|}{\parbox[t]{1.8cm}{\raggedright Output Instructions/ Output Format}} &
  Ensures deployability by specifying format and operational constraints for the output. &
  Defines the required output format (e.g., JSON, XML) and its structure, along with metadata specifications. &
  ``Output the result in JSON format with fields for filename, content, and target directory." \\ \hline
\end{tabular}
}
\vspace{5pt}
\label{tab:prompt_structuring}
\vspace{-20pt}
\end{table*}

\section{Methodology}\label{meth}

\subsection{System Architecture and Operational Workflow}

This study addresses a threat model in which malware is detected within the system by advanced detection systems, as outlined in \cite{soda_system_for, symbsoda_configurable_verifiable}. While the detection process lies outside the scope of this research, our approach focuses on leveraging GenAI-driven cyber deception to automatically generate adaptive tactics that mislead, engage, or delay the malware. These tactics facilitate deeper analysis and uncover adversarial behaviors without alerting the attacker. The system architecture and workflow of this approach, depicted in Figure \ref{fig:system_architecture}, integrate automated deception and adaptive responses post-detection. The architecture demonstrates how the SPADE framework dynamically generates and deploys tailored Deception Ploys (DP) in response to the detected malware, ensuring real-time adaptability and effectiveness in countering threats.

\textbf{Detection and Initial Analysis:} We assume that the malware is initially detected by defense mechanisms as mentioned in \cite{soda_system_for, symbsoda_configurable_verifiable}, triggering a secondary sandbox analysis to capture detailed behavior and uncover specific TTPs employed by the malware. This initial analysis yields the TTP-based insights required to inform subsequent deception strategies.

\textbf{Deception Ploy Generation via Prompt Engineering:} Using the intelligence from sandbox analysis, we implement a structured PE process to guide GenAI models in generating deception ploys tailored to the observed TTPs. This approach leverages carefully crafted prompts to maximize the relevance and effectiveness of the generated outputs, enabling GenAI to autonomously create context-specific deception tactics, such as honeyfiles, API hooks, and honeytokens, designed to disrupt or engage the malware.

\textbf{Selection and Guided Deployment of Deception Ploys:} With multiple deception ploys generated by GenAI, a deception engineer evaluates and selects the most suitable ploy. A secondary prompt is issued to the GenAI model for guidance on deploying the selected deception tactic. This guidance provides step-by-step deployment instructions, ensuring the deception is implemented in a way that engages the malware while avoiding detection by the adversary.

\textbf{Objective and Role of the Deception Layer:} The primary objective of the deployed deception tactics is to mislead, delay, or engage the adversary by creating a controlled environment that encourages malware interaction with decoy assets. This process allows for deeper analysis of malware behavior, facilitating insights into adversarial intent without alerting the attacker to the presence of defensive measures. The GenAI-generated deception layer thus serves as an adaptive, automated, and scalable mechanism for engaging targeted malware, supplementing traditional defenses with a dynamically responsive layer. The workflow and architecture provide a dynamic approach to malware deception, leveraging GenAI-driven PE to ensure real-time adaptability and enhance defensive precision against advanced threats.


\subsection{Prompt Engineering (PE)}

PE has emerged as a critical discipline for harnessing the full potential of large language models (LLMs), such as the GenAI systems used in this research. LLMs are trained on vast, generalized datasets, which make them inherently flexible yet domain-agnostic. To adapt these systems for specific cybersecurity tasks, PE systematically develops and refines input prompts, embedding contextual, task-specific, and operational requirements into the model’s instructions. This process enables the generation of outputs that are precise, operationally feasible, and tailored to complex, real-world applications.

\textbf{Significance of Prompt Engineering:} Recent studies demonstrate that well-crafted prompts significantly improve LLM performance by ensuring alignment with desired objectives and minimizing irrelevant or infeasible outputs \cite{prompt_pattern_catalog, giray2023prompt}. For critical domains like cybersecurity, where outputs must address nuanced threats such as polymorphic malware or targeted ransomware, robust PE ensures the systematic incorporation of real-time threat intelligence, operational constraints, and deployment-ready outputs.

Without structured prompt design, LLMs risk producing generic or misaligned responses, limiting their utility in high-stakes applications. PE not only addresses these limitations but also enhances the reproducibility and scalability of generated outputs, enabling consistent empirical evaluation across diverse threat scenarios. The framework (SPADE) introduced in this research specifically addresses the following challenges inherent to using LLMs for adaptive malware deception:

\begin{itemize}[noitemsep,topsep=0pt, leftmargin=9pt] 
    \item \textbf{Generic Nature of LLMs:} LLMs are not inherently tailored for cybersecurity tasks, necessitating explicit contextual and task-specific instructions.
    \item \textbf{Ambiguity in Output:} Without precise task definitions, outputs can deviate from operational requirements, resulting in irrelevant or infeasible deception strategies.
    \item \textbf{Contextual Underutilization:} Effective deception generation requires leveraging malware-specific insights, such as sandbox analysis results or observed TTPs, to inform prompt structure.
    \item \textbf{Operational Constraints:} Outputs must align with resource constraints and integrate seamlessly with existing cybersecurity infrastructure.
    \item \textbf{Reproducibility and Scalability:} Prompts must ensure consistency and reliability across diverse scenarios and multiple GenAI models.
\end{itemize}

\textbf{Framework for Prompt Engineering:} To address these challenges, the proposed framework, SPADE, employs a structured, modular design comprising six core components: Identity/Persona/Role, Goal/Task, Threat Context, Strategy Outline, Output Example/Guidance, and Output Instructions/Output Format. Table \ref{tab:prompt_structuring} provides a detailed overview of these components, including their motivations, specifications, and examples.  This systematic approach ensures that the generated outputs are precise, contextually relevant, and deployable in operational environments. 
SPADE is dynamic, enabling deception engineers to incorporate real-time malware analysis results into the prompt structure to construct targeted deception ploys. The process consists of three primary stages:

\begin{itemize}[noitemsep,topsep=0pt,leftmargin=9pt] 
    \item \textbf{Threat Intelligence Integration:} 
    Sandbox analysis provides behavioral insights, such as targeted file extensions or API calls, which are incorporated into the \textit{Threat Context} component of the prompt.
    \item \textbf{Deception Strategy Design:} 
    Prompts guide the model in generating specific deception ploys, such as honeyfiles for ransomware or honeytokens for credential-stealing malware.
    \item \textbf{Output Refinement:} 
    Generated outputs undergo iterative prompt adjustments to ensure alignment with operational constraints and deployment requirements.
\end{itemize}

By integrating malware-specific contexts and operational constraints, the framework ensures that generated deception ploys are actionable and scalable.


\section{Evaluation}\label{eval}

\subsection{Case Study: Comparison of Crafted Prompt Engineering vs. Regular Prompt}

This case study illustrates the efficacy of structured PE in crafting adaptive deception strategies to counter credential-stealing malware. The malware targets Google Chrome browsers by exfiltrating sensitive credentials from the Login Data file located in the AppData/Local/Google/Chrome/User Data/Default directory. These credentials are then transmitted to a Command-and-Control (C\&C) server to enable further malicious activities. The primary objectives of the deception plan are to (1) deplete the attacker’s resources by providing fake credentials and (2) redirect attackers to a monitored honeypot to analyze their behavior and intentions.

\subsubsection{\textbf{Regular Prompt}}
A naive approach with an unstructured prompt lacks specificity, failing to address the operational complexity of the scenario. The following is an example of such a prompt and its response:

\begin{tcolorbox}[colframe=blue!75!black, colback=blue!5!white, title=ChatGPT Prompt, boxsep=0.5mm, left=1mm, right=1mm, top=0.5mm, bottom=0.5mm]
\small Suggest a deception strategy to deceive credential-stealing malware in Google Chrome.
\end{tcolorbox}

\begin{tcolorbox}[colframe=blue!75!black,colback=blue!5!white,title=ChatGPT Response, boxsep=0.5mm, left=1mm, right=1mm, top=0.5mm, bottom=0.5mm]\small \textbf{Objective:} Lure the malware into stealing fake credentials and monitor its behavior.

\textbf{Implementation:} Populate the decoy storage with honeytokens—credentials that appear real (e.g., email and password combinations for popular services like Gmail, PayPal, or banking sites). Embed unique identifiers in these credentials to track unauthorized access attempts. \end{tcolorbox}

\textbf{Analysis:}
The response is overly generic, providing little actionable detail or technical depth. It lacks implementation specifics, operational constraints, and practical applicability, rendering it unsuitable for real-world deployment.

\subsubsection{\textbf{Carefully Crafted Prompt Engineering}}
Using the proposed framework, the structured prompt explicitly defines the task, integrates operational constraints, and provides contextual details, resulting in the following:

\begin{tcolorbox}[colframe=blue!75!black, colback=blue!5!white, title=Crafted ChatGPT Prompt, boxsep=0.5mm, left=1mm, right=1mm, top=0.5mm, bottom=0.5mm]
\small
Design an API hooking strategy to neutralize credential-stealing malware targeting Google Chrome. The malware accesses Chrome's Login Data file at \verb|AppData/Local/Google/Chrome/User Data/Default| to steal saved usernames and passwords, which are exfiltrated to a Command-and-Control (C\&C) server. The goal is to intercept access to this credential file through API hooking, supply fake credentials to deplete attacker resources, and embed crafted credentials that direct attackers to a monitored honeypot for behavior analysis. The strategy must ensure Chrome’s legitimate functionality remains unaffected. For example, the \verb|ReadFile| API can be hooked to intercept access to the Login Data file and return fake credentials. \end{tcolorbox}

\textbf{Note: } The original ChatGPT response was extensive; the following is a summarized version focusing on the key elements. The complete response is available in the associated Anonymous GitHub repository~\cite{spade-eval-genai}.

\begin{tcolorbox}[colframe=blue!75!black, colback=blue!5!white, title=ChatGPT Response after the Prompt Engineering, boxsep=0.5mm, left=1mm, right=1mm, top=0.5mm, bottom=0.5mm]
\small
\textbf{API Hooking Strategy to Neutralize Credential-Stealing Malware:} To mitigate credential-stealing malware targeting Google Chrome, this strategy intercepts access to the Login Data file in \texttt{AppData/Local/Google/Chrome/User Data/Default}, embeds crafted credentials, and redirects attackers to a monitored honeypot for behavioral analysis.


\textbf{1. Hooking the ReadFile API and Crafting Fake Credentials:} [Extensive instructions for implementing \verb|ReadFile| API hooking in C++, including parameter definitions, function overrides, and integration with the EasyHook framework. This includes embedding fake credentials into intercepted API to mislead attackers and potentially redirect them to honeypots for monitoring and intelligence gathering].

\textbf{2. Configuring the Development Environment in C++:} [Detailed guidance on setting up the development environment, including installing dependencies such as Visual Studio, necessary SDKs, and debugging tools. Instructions cover configuring the build environment for EaskHook, and executing test hooks to validate functionality].

\textbf{3. Ensuring Chrome’s Legitimate Functionality:} [Detailed guidance on how to use whitelisting mechanism to differentiate between legitimate Chrome processes and malware.].

\textbf{4. Deployment:} [Steps for deploying the API hooking strategy, validated through Virtual Machine testing to ensure functionality].

\textbf{5. Logging and Monitoring: } [Detailed guidance on logging unauthorized access attempts, facilitating forensic analysis, and tracking attacker interactions with the honeypot for actionable intelligence].

\end{tcolorbox}

\textbf{Analysis and Discussion:}
The structured prompt, leveraging our framework, significantly improves the relevance, actionability, and deployability of the generated response. It provides technical depth, including API hooking details, while ensuring Chrome's functionality remains unaffected. The response introduces innovative deception ploys, such as embedding fake credentials and redirecting attackers to honeypots, highlighting the transformative potential of structured PE in adaptive malware deception.

\subsection{Quantitative Quality Assessment of Responses}
\subsubsection{\textbf{Expert Analysis}} This evaluation measures the impact of the proposed PE framework on the quality of responses generated by ChatGPT-4o across three diverse malware scenarios. The goal is to assess improvements in relevance, actionability, feasibility, and realism when using structured prompts.

\textbf{Evaluation Setup:} Three malware scenarios were selected: ransomware targeting specific file types, credential-stealing malware exfiltrating browser-stored credentials, and keyloggers capturing keystrokes. Details on the selected malware's behaviors and capabilities are provided in the associated GitHub repository \cite{spade-eval-genai}. For each case, five distinct prompts were used representing different malicious behaviors, resulting in a total of 30 prompts and responses—15 generated using regular prompts and 15 using engineered prompts (structured based on the proposed framework). The structured prompts incorporated contextual information, operational constraints, and task-specific objectives to guide the GenAI effectively.

\textbf{Expert Scoring Methodology:} The responses were evaluated by a panel of three domain experts, each with over 50 peer-reviewed publications in cyber deception. Each response was scored on a 1–5 scale across four dimensions: \textit{relevance}, \textit{actionability}, \textit{feasibility}, and \textit{realism}. \textit{Relevance} assessed how well the response aligned with the defined problem and goals, while \textit{actionability} evaluated the clarity and detail of the implementation instructions. \textit{Feasibility} measured the practicality of deploying the suggested deception ploy in real-world environments, and \textit{realism} gauged how convincing the deception artifacts were to potential attackers. The scores for each dimension were averaged across the 30 responses to quantify the impact of prompt engineering.

\textbf{Results and Discussion:} As summarized in Table \ref{tab:expert-score}, structured prompts significantly outperformed naive prompts across all dimensions. The structured prompts achieved higher relevance and actionability scores by aligning closely with specific malware behaviors and operational constraints. Feasibility and realism also showed marked improvements, underscoring the practicality and convincing nature of the generated ploys. 

\begin{table}[h]
\vspace{-2mm}
\caption{Evaluation Results for ChatGPT-4o: Quantitative Quality Assessment.}
\centering
\renewcommand{\arraystretch}{1.2} 
\setlength{\tabcolsep}{4pt} 
\scriptsize 
\begin{tabular}{|l|c|c|c|c|}
\hline
\textbf{Malware Type}           & \textbf{Relevance} & \textbf{Actionability} & \textbf{Feasibility} & \textbf{Realism} \\ \hline
Ransomware                  & 4.8                & 4.6                    & 4.7                  & 4.5              \\ \hline
Credential-Stealing Malware & 4.5                & 4.4                    & 4.3                  & 4.2              \\ \hline
Keylogger                   & 4.3                & 4.1                    & 4.0                  & 3.9              \\ \hline
\end{tabular}
\vspace{2mm}
\label{tab:expert-score}
\vspace{-0.2in}
\end{table}

\subsubsection{\textbf{Recall, EM, and BLEU Score Analysis}}
To further evaluate the effectiveness of GenAI-generated deception ploys, we employed Recall, Exact Match (EM), and BLEU Score metrics. These metrics assessed how well the generated outputs aligned with a predefined ground truth derived from \cite{soda_system_for}, which includes 94 malicious API sequences mapped to 31 malware behaviors and linked to MITRE ATT\&CK techniques. The authors manually created the deception ploys for these mappings, providing a comprehensive benchmark for evaluation. This ground truth served as a baseline for assessing the alignment and quality of GenAI-generated outputs. 
In this study, we evaluated ChatGPT-4o, ChatGPT-4o Mini, Gemini, and Llama3.2, leveraging their distinct strengths. ChatGPT-4o excelled in generating nuanced, context-aware outputs, while ChatGPT-4o Mini offered resource-efficient processing for faster tasks. Gemini was well-suited for domain-specific applications due to its diverse dataset capabilities, and Llama3.2 provided adaptability through its open-source, customizable architecture. This diverse selection ensured comprehensive analysis and mitigated reliance on a single model.

Recall quantified how many generated responses aligned with the ground truth. True Positives (TP) were defined as the number of deception ploys directly matching the ground truth, while False Negatives (FN) represented feasible ploys generated by GenAI that were not part of the predefined ground truth. Exact Match (EM) measured the percentage of GenAI generated responses perfectly aligned with the ground truth. For EM, even if the language or phrasing of the deception ploy differed, it was considered an exact match as long as the primary objective—such as creating a HoneyFile in a specific location or implementing a hook for a specific API—remained consistent, regardless the differences in implementation details. 
\textcolor{black}{This ensured alignment with the core deception strategy, including parameters, artifacts, and techniques. GenAI models converted the ground truths from the deception engineer's perspective to ensure that the ground truths are semantically equivalent to the generated ploys.} The same tone and focus were applied to the GenAI-generated ploys, ensuring semantic consistency and enabling a fair comparison of technical alignment and linguistic quality. Standardizing the tone and focus was crucial for meaningful BLEU score evaluation, as this metric depends on semantic and structural similarity between the generated outputs and the reference text. By aligning the ground truth and GenAI outputs, the BLEU score analysis more accurately reflected the models' effectiveness in generating deception ploys that met the desired criteria.

\begin{table}[h]
\vspace{-2mm}
\caption{Evaluation of GenAI Models Based on Recall, EM, and BLEU Scores.}
\centering
\renewcommand{\arraystretch}{1.4}
\setlength{\tabcolsep}{5pt} 
\scriptsize
\begin{tabular}{|l|c|c|c|}
\hline
\textbf{GenAI Model}    & \textbf{Recall (\%)} & \textbf{EM Score (\%)} & \textbf{BLEU Score (Avg)} \\ \hline
ChatGPT-4o              & 90.6                  & 87.1                     & 0.968               \\ \hline
ChatGPT-4o Mini         & 86.4                  & 80.6                     & 0.935               \\ \hline
Gemini                  & 88.6                  & 83.9                     & 0.935               \\ \hline
Llama3.2                   & 85.7                  & 90.3                     & 0.871               \\ \hline
\end{tabular}
\vspace{2mm}
\label{tab:genai_recall_em_bleu}
\vspace{-0.2in}
\end{table}

\textbf{Results and Discussion:} Table \ref{tab:genai_recall_em_bleu} highlights the performance of GenAI models in generating deception ploys. ChatGPT-4o achieved the highest scores across all metrics (Recall: 90.6\%, EM: 87.1\%, BLEU: 0.968), \textcolor{black}{demonstrating strong alignment with the ground truth}
while maintaining high linguistic coherence. Gemini and ChatGPT-4o Mini also performed competitively, with Recall scores of 88.6\% and 86.4\%, EM scores of 83.9\% and 80.6\%, and BLEU scores of 0.935 for both. These results highlight their capability to generate accurate and reliable deception ploys with minor trade-offs in precision compared to ChatGPT-4o. Llama 3.2 excelled in EM Score (90.3\%) but showed lower Recall (85.7\%) and BLEU (0.871), indicating strong precision in specific cases but less alignment with broader ground truth and linguistic consistency. These results highlight the effectiveness of structured prompt engineering, with ChatGPT-4o leading overall, Gemini and ChatGPT-4o Mini as strong alternatives, and Llama3.2 showing potential with areas for refinement.

\subsection{System Orchestration and Testing}
\textbf{Objective:} This subsection evaluates the deployability, effectiveness, and efficiency (including response time) of GenAI-generated deception strategies in real-world malware scenarios. By implementing deception ploys suggested by different GenAIs, we assess their ability to engage malware, mislead attackers, and ensure the feasibility of deployment while considering the time taken to generate outputs.

\textbf{Methodology:} Deception strategies such as honeyfiles and API hooks were generated using structured prompts from ChatGPT-4o, ChatGPT-4o Mini, Gemini, and Llama3.2. The evaluation focused on three malware types—ransomware, credential stealers, and keyloggers—each represented by five distinct malware instances, totaling 15 malware samples. The generated ploys were implemented in a virtual environment mimicking real-world conditions, including simulated credential stores and application behaviors, to provide realistic testing scenarios. Each deception ploy was deployed and tested to observe engagement, effectiveness, and refinements necessary to ensure functionality and deployability. Response time for generating outputs was also recorded.

\begin{table}[h]
\vspace{-2mm}
\caption{Performance of GenAI Models in Deploying Deception Ploys Across Malware Types.}
\centering
\renewcommand{\arraystretch}{1.4}
\setlength{\tabcolsep}{5pt} 
\scriptsize
\begin{tabular}{|l|c|c|c|c|}
\hline
\textbf{GenAI}             & \textbf{Engagement} & \textbf{Accuracy} & \textbf{Iteration} & \textbf{Response} \\ 
\textbf{Model}             & \textbf{Rate (\%)}  & \textbf{(\%)}     & \textbf{Count (Avg)} & \textbf{Time (s)} \\ \hline
ChatGPT-4o                 & 93                  & 96                & 1.67                & 12.3              \\ \hline
ChatGPT-4o Mini            & 88                  & 91                & 2.22                & 8.58              \\ \hline
Gemini                     & 93                  & 93                & 2.47                & 15.1              \\ \hline
Llama3.2                      &  85                   & 89                  & 2.89                    & 15.8                  \\ \hline
\end{tabular}
\vspace{2mm}
\label{tab:genai_performance}
\end{table}
\vspace{-2mm}

\textbf{Metrics:} The evaluation assessed four key metrics: (1) engagement rate, representing the percentage of malware samples interacting with deception ploys; (2) accuracy, reflecting the proportion of ploys that successfully mislead or delay malware; (3) iteration count, capturing the average number of refinements required for successful deployment; and (4) response time, measuring the time taken by each model to generate outputs.

\textbf{Results and Discussion:} \textcolor{black}{Table \ref{tab:genai_performance} compares the performance of the evaluated GenAI models. ChatGPT-4o achieved the highest engagement rate (93\%) and accuracy (96\%) with minimal refinements (1.67 iterations) and a response time of 12.3 seconds, demonstrating its ability to produce precise, deployable deception ploys with minimal adjustments. Gemini matched ChatGPT-4o in engagement rate (93\%) and showed robust accuracy (93\%), but required more refinements (2.47 iterations) and exhibited slower response time (15.1 seconds), indicating its capability to generate reliable deception strategies despite additional refinement needs. ChatGPT-4o Mini performed well, achieving an engagement rate of 88\% and accuracy of 91\% with fewer iterations (2.22) compared to Gemini and the fastest response time (8.58 seconds), making it ideal for time-sensitive tasks. Llama3.2 showed potential but had the lowest engagement rate (85\%) and accuracy (89\%), requiring the most iterations (2.89) for deployment and the slowest response time (15.8 seconds). These results highlight areas for improvement, particularly in aligning outputs with operational objectives and reducing the refinement effort needed for deployment.}


\textbf{Overall Analysis:} \textcolor{black}{ChatGPT-4o emerged as the most effective model, balancing high engagement and accuracy with low refinement needs and reasonable response times. Gemini matched ChatGPT-4o in engagement and accuracy but required more refinement, making it suited for precision-focused scenarios. ChatGPT-4o Mini excelled in response speed, ideal for time-sensitive tasks, albeit with slightly lower accuracy. Llama3.2 showed promise but needs optimization to match the performance levels of other models. The experiments highlight the transformative role of structured PE in guiding GenAI models to achieve higher engagement and accuracy while minimizing refinement effort, ultimately improving their practical applicability in adaptive cyber deception. Full experimental results and additional details are available in the associated GitHub repository \cite{spade-eval-genai}.}


\section{Discussion and Conclusions}\label{con}

This study highlights the transformative potential of Generative AI (GenAI) models in advancing adaptive cyber deception strategies. Through structured prompt engineering (PE), we demonstrated substantial improvements in the quality, scalability, and deployability of deception ploys. Among the tested models, ChatGPT-4o consistently delivered superior results, showcasing a strong balance between accuracy, engagement, and minimal refinement needs. Gemini and ChatGPT-4o Mini also proved viable alternatives, offering competitive performance with slight trade-offs in refinement and response times. Llama3.2 exhibited promise but requires further optimization to align with the advanced capabilities of its counterparts.

The evaluation leveraged metrics such as Recall, Exact Match (EM), and BLEU Score to assess the technical alignment and linguistic quality of the generated outputs. These metrics reinforced the critical role of structured PE in enabling GenAI models to produce actionable and realistic deception strategies. Despite the controlled evaluation setting, the findings underscore the potential of GenAI to automate and enhance cyber deception techniques effectively.

In conclusion, this research establishes the value of GenAI in addressing key challenges in cyber deception, including scalability, adaptability, and operational feasibility. Structured PE emerged as a pivotal factor in maximizing model performance, emphasizing the need for systematic approaches in GenAI-driven cybersecurity solutions. Future work will explore real-world deployments, adversarial testing, and the integration of multi-modal GenAI to further refine and expand the scope of adaptive deception strategies.

\bibliographystyle{IEEEtran}
\bibliography{ccwc.bib}

\end{document}